\documentclass[aps,prl,preprint]{revtex4}
\usepackage{graphicx}

\begin{document}


\title{On self-protecting singlets in cuprate superconductors}


\author{J. R\"ohler}

\email{abb12@uni-koeln.de}

\homepage{http://www.uni-koeln.de/~abb12}

\affiliation{Universit\"at zu K\"oln, Z\"ulpicher Str. 77, D-50937 K\"oln, Germany}


\date{\today}

\begin{abstract}
The basal area (Cu--Cu grid) of the cuprate superconductors not only
tends to shrink on hole doping, as expected from single electron
quantum chemistry, but exhibits also an electronically incompressible
``hump'' around optimum doping $n_{opt}\simeq 0.16$.  The hump
collapses near critical doping $n_{opt}\simeq 0.19$.  We analyze the
origin of the hump in terms of a classical liquid of interacting 
incompressible particles in a container with antiferromagnetic walls. 
Oxygen holes interacting with the wall form singlets, 
protect themselves against other holes by an incompressible ``spin 
fence'', and thus interact also with the lattice. Occupation of the CuO$_{2}$ lattice
with holes must therefore follow a non-double-occupant constraint
also for the oxygen cage enclosing the copper hole. 
Closest packing of self-protecting singlets is found to occur around 
critical doping; closest packing of $paired$ self-protecting singlets around 
optimum doping.  These singlet-states are bosonic, but are not magnetic polarons.

\end{abstract}

\pacs{}
\keywords{superconductivity, cuprates, lattice effects}

\maketitle


\section{\label{Intro}Introduction}

The lattice parameters of the cuprate superconductors are well known
strong functions of doping.  It is however only poorly understood
whether their doping dependence is extrinsically controlled from
the complex stereochemistry of the doping blocks outside, or
intrinsically by the properties of the strongly correlated
electron states residing inside the CuO$_2$ planes.  The external chemistry and
the embedded metallic layer reside in the same unit
cell, and hence both are expected to co-determine the doping induced
variations of the cell parameters.  How to disentangle the most
important intrinsic effects from the extrinsic ones?  Detailed
inspection of many of the available lattice data 
\cite{Kal01, RadJor, FukTan, BoeFau, HuaChu, JorMit}, displaying doping
dependencies in the single- and multi-layer compounds of different
families, shows forces of different nature to govern the behaviours of the 
$a$\/-, $b$\/-parameters in the planar directions, and of the $c$\/- parameter 
in the perpendicular direction. $a$ and $b$ tend to
contract with increasing hole concentration, almost independent on the
number of CuO$_2$ layers and the chemistry of the doping block. 
Uniaxial strain from the dopants may mask or even invert the
shortening of one of the basal bondlengths, $e.g.$ in underdoped
YBa$_2$Cu$_{3}$O$_{x}$ where $a$ contracts as $b$ expands
\cite{Kal01}. But the basal area $B$, defined by the square of the
basal Cu--Cu distances, turns out almost unaffected by strain from the
doping block.  Thus $B$ allows better for comparisons between
compounds from different families than the individual basal bond
lengths.  The $c$\/-parameter however seems to behave arbitrarily: as
the basal plane contracts, $c$ may expand, contract, or both
\cite{RadJor, HuaChu, Kal01, FukTan}. Apparently the problem of 
entangled intrinsic and extrinsic doping effects on the 
lattice seems to solve quite naturally: the behaviour of the basal plane is 
governed by the nearly-2D quantum liquid residing in it, while that 
of the perpendicular parameter is dominated by the chemistry of the environment. 
In this article we will focus on the doping dependence of the basal 
plane $B(x)$, in particular on the origin of the strong hump around 
optimum doping, see Figs.~\ref{fig:LSCO},~\ref{fig:HBCO1L},~\ref{fig:YBCO},
~\ref{fig:HBCCO2L}.

\section{\label{experiment}Experimental data}

\begin{figure}
\includegraphics*[width=15cm]{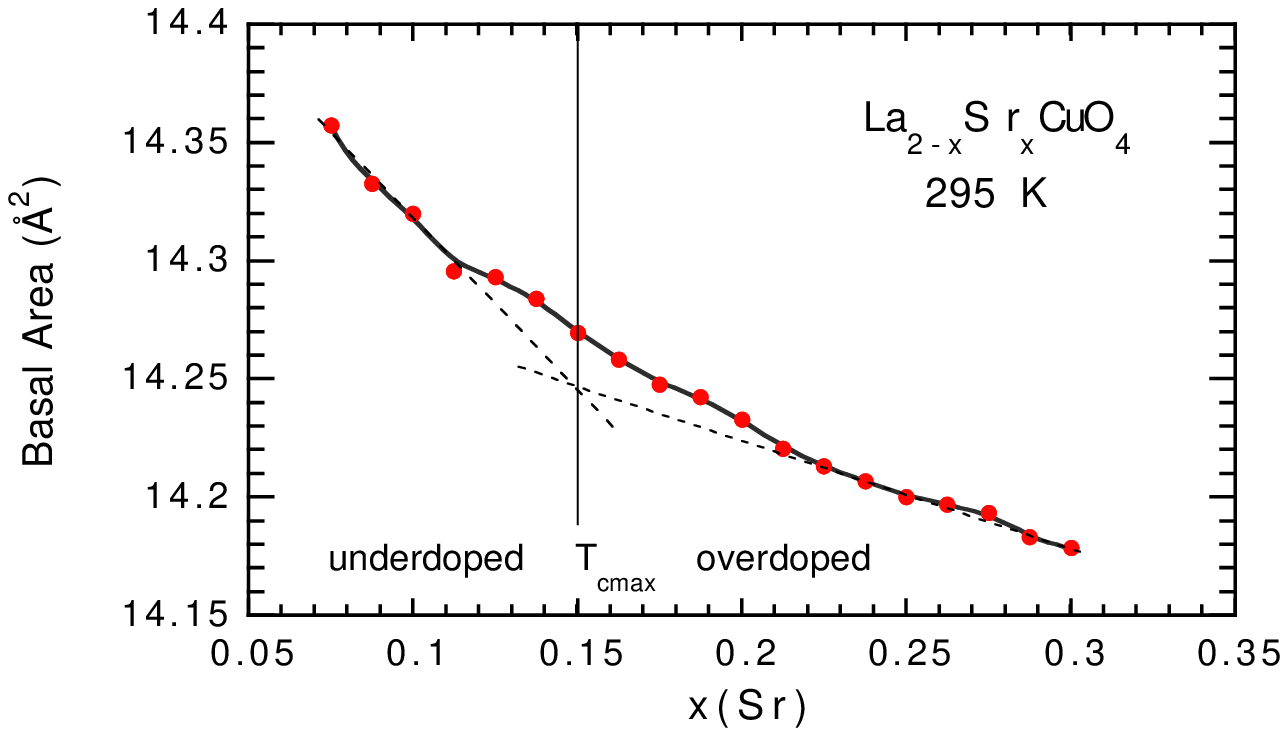}
\caption {\label{fig:LSCO}Basal area of one-layer 
La$_{2-x}$Sr$_{x}$CuO$_{4}$ as a function of doping using 
neutron diffraction data of Radaelli et $al.$ \cite{RadJor}.}
\end{figure}

\begin{figure}[b]
\includegraphics*[width=13.35cm]{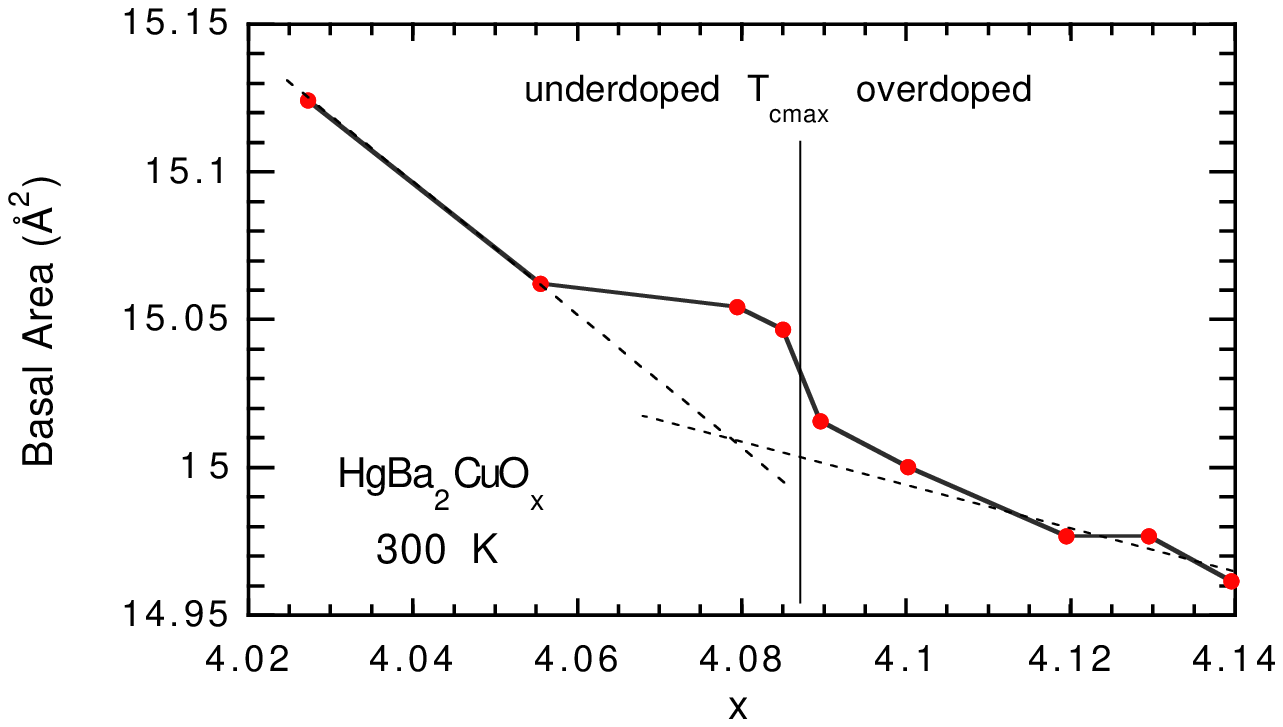}
\caption {\label{fig:HBCO1L}Basal area of one-layer HgBa$_{2}$CuO$_{x}$ 
as a function of doping using x-ray diffraction data of Fukuoka et $al.$  \cite{FukTan}}.
\end{figure}

\begin{figure}[t]
\includegraphics*[width=15cm]{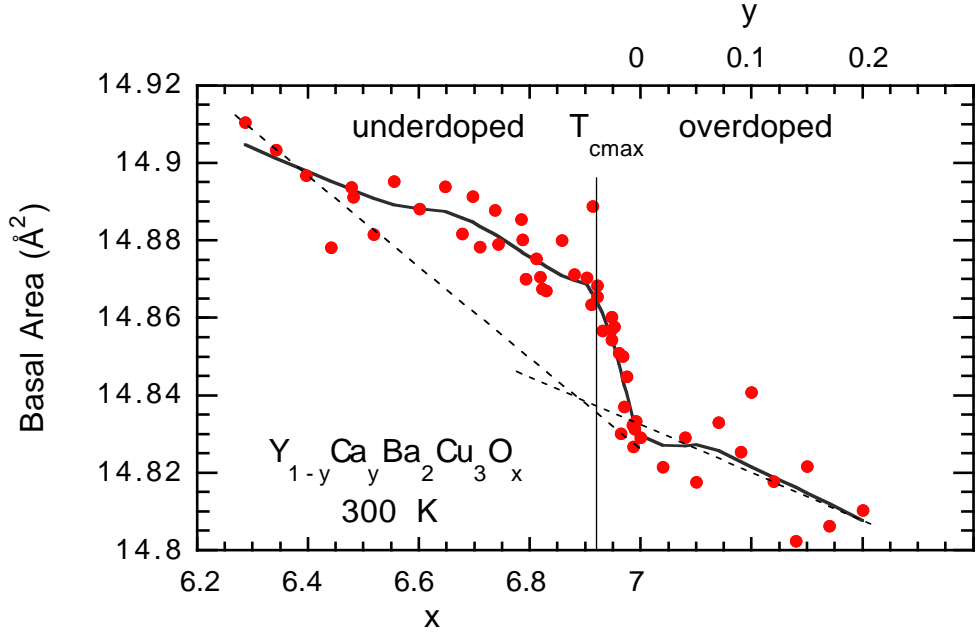}
\caption{\label{fig:YBCO}Basal area of two-layer 
Y$_{1-y}$Ca$_{y}$Ba$_{2}$Cu$_{3}$O$_{x}$ 
using x-ray diffraction data of B{\"o}ttger et 
$al.$ \cite{BoeFau} (overdoped by Ca, $y\simeq 0.96=const.$ ), and Kaldis 
and collaborators \cite{Kal01} (underdoped).}
\end{figure}

\begin{figure}[b]
\includegraphics*[width=15cm]{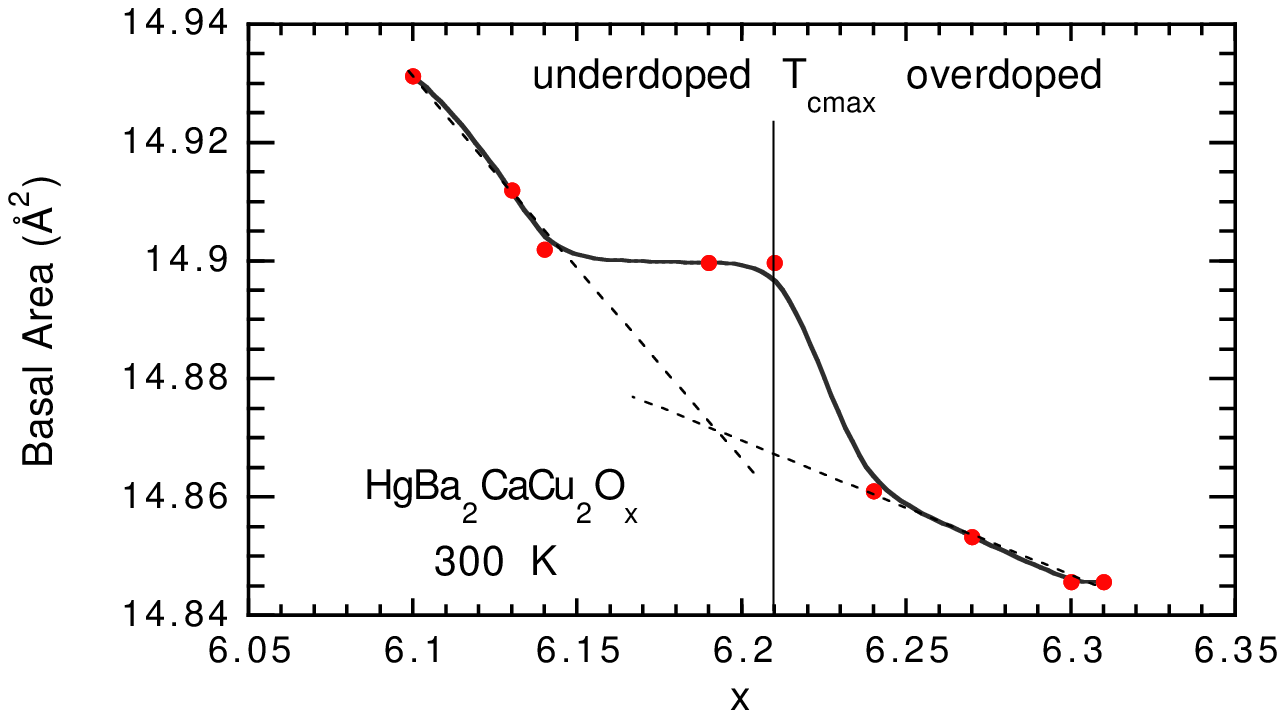}
\caption{\label{fig:HBCCO2L}Basal area of two-layer HgBa$_{2}$CaCu$_2$O$_{x}$ 
as a function of doping using x-ray diffraction data of Fukuoka et $al.$  \cite{FukTan}}
\end{figure}

Figs.~\ref{fig:LSCO},~\ref{fig:HBCO1L},~\ref{fig:YBCO},~\ref{fig:HBCCO2L} 
display the basal areas $B$ of typical one- and 
two-layer cuprates as a function of doping. 
$B$ refers to the grid of $nn$ planar copper atoms as reported from 
x-ray or neutron diffraction measurements at room-temperature by Radaelli 
et $al.$ \cite{RadJor} (La$_{2-x}$Sr$_{x}$CuO$_{4}$, $T_{cmax}=36$ K), 
Fukuoka et $al.$  \cite{FukTan} (HgBa$_{2}$CuO$_{x}$, $T_{cmax}=96$ K; 
and HgBa$_{2}$CaCu$_{2}$O$_{x}$, $T_{cmax}=127$ K) , B{\"o}ttger et 
$al.$ \cite{BoeFau}, and Kaldis \cite{Kal01} 
(Y$_{1-y}$Ca$_{y}$Ba$_{2}$Cu$_{3}$O$_{x}$, $T_{cmax}=92$ K).
The thick drawn out lines connecting the data points are guides to the eye.

The overall behavior of $B(x)$ exhibits surprisingly strong
similarities in all systems under comparison: $i.$ As expected from
the increasing covalency with hole doping the basal areas shrink
between the insulator--metal transition and the strongly overdoped
regime.  The contraction is in the order of 1.5\% $ii.$ $B(x)$
exhibits a hump centered around optimum doping.  The hump is
weakest in La$_{2-x}$Sr$_{x}$CuO$_{4}$, and strongest in
HgBa$_{2}$CaCu$_{2}$O$_{x}$.  Its maximum is centered aroud optimum
doping $x_{opt}$.

Following quantum chemical approximations hole doping will remove 
electrons from the antibonding $\sigma^*$Cu$3d_{x^2-y^2}$O$2p_{x,y}$ 
band, increase the amount of 
covalent character in the Cu--O bonds, and will shorten
them.  The hump indicates a significant deviation from this 
one electron bandstructure picture.  The thin dashed straight lines in
Figs.~\ref{fig:LSCO},~\ref{fig:HBCO1L},~\ref{fig:YBCO},~\ref{fig:HBCCO2L} 
are fitted to the data points at
the strongly under- and overdoped ends and, extrapolated towards the
underdoped-overdoped phase boundary, turn out to intersect around
$x_{opt}$.  They may serve as coarse approximations for the quantum
chemical ``background'' $B_{0}(x)$.  
Its change of slope around $x_{opt}$ might indicate that the Cu--O
bonding changes from semicovalency in the underdoped to covalency in
the overdoped regime.

Evidently the hump stems from the strong correlations of the holes in
the CuO$_2$ planes.  Its location at optimum doping points also to a
connection with the superconductivity occuring at much lower
temperatures than 300 K. Consider the compressibility of the quantum
liquid in the planes is approximated by $\kappa_{e}\propto -\partial
(B-B_{0})/\partial n$, where $n$ is the number of holes/Cu/unit area, and
$n\propto x$.  Then $\kappa_{e}\simeq 0$ close $x_{opt}$, $i.e$ the
quantum liquid is incompressible around the optimum hole concentration
$n_{opt}=0.15-0.16$, notably not at quarter filling $n=0.25$.
 
In some compounds the collapse of the hump 
occurs together with subtle structural instabilities in the crystallographic 
cell, of martensitic type in YBa$_{2}$Cu$_{3}$O$_{x}$ 
\cite{KalLoe,Roe02a}, or of order-disorder type in 
HgBa$_{2}$Cu$_{2}$O$_{x}$ \cite{JorMit}. Evidently these lattice 
instabilities are connected with the transition into the overdoped 
regime, but are most likely not at its origin. 

Stable long range ordered nano domains might be another source of the hump. 
La$_{2-x}$Sr$_{x}$CuO$_{4}$, due to its octahedral tilts $the$ system most 
susceptible to stable nano domains, exhibits however the weakest hump. 
On the other hand the Hg-cuprates with their almost flat CuO$_2$ 
lattices, devoid of structural compliance, exhibit the strongest humps.

\section{\label{model}The model}

Consider $B(x)$ reproduces the equation of state of the quantum liquid
confined to the CuO$_2$ lattice.  Suppose that the temperature is
sufficiently high and the density of classical particles sufficiently 
low, then $B(x)$ exhibits a striking similiarity with the van der Waals
equation, or related types describing classical real gases. Here we
describe the particles by singlets of oxygen hole and copper spins. 
We address the question, how a 
quantum liquid, created through singlet formation in a hole doped 
CuO$_2$ lattice, may be connected with lattice properties such as 
$B(x)$ and its hump around $x_{opt}$. Further down we 
will also try to justify why the problem 
may be phrased in terms of a classical Bose gas scheme.

We consider the antiferromagnetic lattice of the copper spins  
as the walls of the ``container'' enclosing the available area for the 
spin singlets created upon hole doping. The moving holes exert 
a mean pressure (positive or negative) on the antiferromagnetic lattice given by 
$p\propto nt$ where $t$ is the average kinetic energy of the 
hole. Note that a
Fermi liquid confined in the volume of its metal exerts its mean 
pressure only on the atomic lattice. 

A doped hole at an oxygen site destroys the antiferromagnetic order in
its vicinity and thereby leads to an attractive interaction when it
shares the region of depressed antiferromagnetic order with a copper
hole.  The attractive interaction creating the spin singlet thus tends
to keep the oxygen and Cu atoms closer together than it would be the
case for noninteracting holes.  B{\"ottger} and Dichtel \cite{BoeDic}
find from a three-band Hubbard model an oxygen displacement per hole
by -$0.04$ {\AA} along the Cu--O bond.  The creation of singlets has
thus the same effect as a slight compression of the basal area
complying with an increasing pressure.  Thus the oxygen holes interact
also strongly with the atomic lattice ``eating'' spins in the antiferromagnetic
wall of its container. 

On the other hand short-range repulsive forces between the singlets 
keep them sufficiently apart to prevent them from
occupying the same places at the same time. The area occupied by a 
singlet themself must be thus subtracted from the area available to
any other singlet in the container. For sufficiently high doping 
the repulsive interaction between singlets will  
start to outweigh the contraction of lattice driven by singlet 
creation. Long range attractive interactions between the singlets 
will pack them more closely in the container and thus may act as 
an additional pressure.

The potential energy $U$ of the interaction between the 
singlets may be expressed only by their relative separation $R$, in its simplest form as 

\begin{eqnarray}
U(R)=\left\{\begin{array}{l@{\quad:\quad}l} \infty &  R<R_{0}
\\ -U_{0}(R_{0}/R)^s  &  R>R_{0}  \end{array}\right.
\label{eq:one}
\end{eqnarray}

$R_{0}$ is the minimum possible separation between incompressible 
singlets. The exponent $s$ is $e.g. \simeq 6$ in typical van der Waals gases.

Long range interactive forces may create also 
paired singlets excluding a larger area than two unpaired singlets. 
Then the area available to other singlets will be even more reduced 
than by closely packed unpaired singlets. As a result 
formation of paired singlets will render the CuO$_2$ lattice 
even less compressible than most closely packed unpaired singlets. 
It is therefore suggesting to assume that the optimum 
doped incompressible CuO$_2$ lattice accomodates its holes ($n_{opt}= 0.16$) in
closest packed $paired$ singlets.

\subsection{The self-protecting singlet (SPS)}

How large will be the incompressible areas covered by a paired
singlet, and a unpaired singlet, respectively ?  $Paired$ singlets in
the most closely packed conformation have to match the maxium of the
hump at $n_{opt}\simeq 0.16$.  $Unpaired$ singlets in the most closely
packed conformation are expected to match a critical hole
concentration $n_{crit}$, slightly larger than $n_{opt}$.  For
$n>n_{opt}$ paired singlets will start to overlap with each other 
and thus will be broken, and the hump will collapse.  
We locate the collapse of the hump
around $n_{crit}\simeq 0.19$. For $n>n_{crit}$ even 
unpaired singlets will be broken, but may create a new
type of singlet state excluding a smaller area than the singlets 
at $n<n_{cit}$.

We consider a perfect antiferromagnetic CuO$_2$ lattice. The
non-double-occupant constraint for the Cu$3d^9$ sites requires that doping
must create additional holes on the oxygen sites.  Hirsch \cite{Hir}
showed that without flipping Cu spins a doped oxygen hole can only
move within a ``cage'' of four oxygen atoms surrounding the nearest Cu
atom.  This is due to the phase coherence in the symmetric combination
of the four oxygen states.   Zhang and Rice\cite {ZhaRic} worked out that the binding
energy of the resulting spin-singlet state of  the symmetric oxygen 
hole and the Cu hole is 4 times higher compared to that of a
spin-singlet state of an oxygen hole sitting at a fixed site, and the
Cu hole. Thus the so-called Zhang-Rice (ZR) singlet distinguishes 
itself from other possible singlet states by its extraordinarily high stability. 
The cage determines the area covered by a ZR singlet and is  
given by $d_{O-O}^2=a^2/2$. Here $d_{O-O}$ is the $nn$ 
oxygen-oxygen distance, and $a$ the Cu--Cu distance.  
The ZR singlet has no magnetic interactions with
all other copper holes. But two neighbored holes trying to share their oxygen
cages with each other will feel a strong repulsion.  Thus formation of 
$nn$ singlets in connected cages will be very unlikely. 
Rather the ZR singlets will strongly repel each
other, and thus will create an excluded area around their cages.

To protect the symmetric oxygen hole in the cage from other holes, the
excluded area has to extend at least over the four neighbored cages
containing the 4 $nn$ Cu spins. The 4 $nn$ Cu spins are 
exempted from the formation of own singlet states but act as a ``fence'' 
enclosing the ZR singlet. We label this as ``self-protection'' 
of the ZR singlet for a safe life, at least within its spin fence. 
Hence a Cu$_5$O$_{16}$ cluster occupied by one oxygen hole in the 
central cage appears to be the minimum conformation for a stable 
singlet, that is a self protecting singlet (SPS). It covers nine $nn$ oxygen-oxygen 
squares with an area $3d_{O-O}^2 = 9/2 a^2$.  Fig. \ref{sps} 
displays a SPS in the antiferromagnetic CuO$_2$ lattice. The circle 
enclosing the central cage indicates the ZR singlet, the hatched 
cages the excluded area with the spin fence.

The Cu$_5$O$_{16}$ cluster comprising a self protecting singlet (SPS)
looks alike a polaron, but is not a conventional spin polaron
polarizing the Cu lattice ferromagnetically. Here holes and spins are
in the same band unlike the situation in the magnetic polarons.

Upon propagation of the SPS from a given site to a neighbored site the
oxygen hole has to flip its spin, and will thus be able to create a
singlet with the $nn$ Cu hole.  The Cu spin of the abandoned site
will recover and hence a possible ``loop-hole'' in the spin fence will be
closed.  Thus the spin fence moves together with the propagating hole while the
antiferromagnetic lattice remains intact.  This scenario has
appealing similarities with the ``spin-bag'' mechanism proposed by
Schrieffer $et$ $al.$ \cite{SchZha} in that both, the spin-bag and the
self protecting singlet, are both polaron-like, but are not
conventional spin-polarons with heavy masses.  We understand however
that a fundamental difference occurs in that in the spin-bag
mechanism the propagating hole and its surrounding bag act as
fermionic quasiparticle, wheras a hole propagating in a self protecting
singlet has to be considered as a bosonic particle.

We may phrase the description of the SPS in terms of another
non-double-occupant constraint for the doped holes: the grid of the 5
oxygen corner linked oxygen cages in a Cu$_5$O$_{16}$ cluster may
be only singly occupied.  Or alternatively expressed: the spin fences
of self-protecting singlets are not allowed to overlap.

Respecting both non-double-occupant constraints,
that for the Cu sites and that for the oxygen cages, it is apparent 
that the maximum number of singlet states in a doubly constrained 
CuO$_2$ lattice will be 
much smaller than in the only singly constrained case. 
We will show further below that the doubly constrained doping of 
holes in a CuO$_2$ lattice will be generically inhomogenous for sufficiently 
high concentrations. Ignoring the non-double-occupant constraint for the oxygen cage 
will however allow for homogenous hole distributions at optimum 
doping, as $e.g.$ in the RVB models \cite{And97}.

Using the information from the experimental $B(x)$ we will
estimate the minimal size of the particles occupying the CuO$_2$
lattice in the different regimes of doping.

\subsection{The paired self-protecting singlet (PSPS)}

\begin{figure}
\includegraphics*[width=7cm]{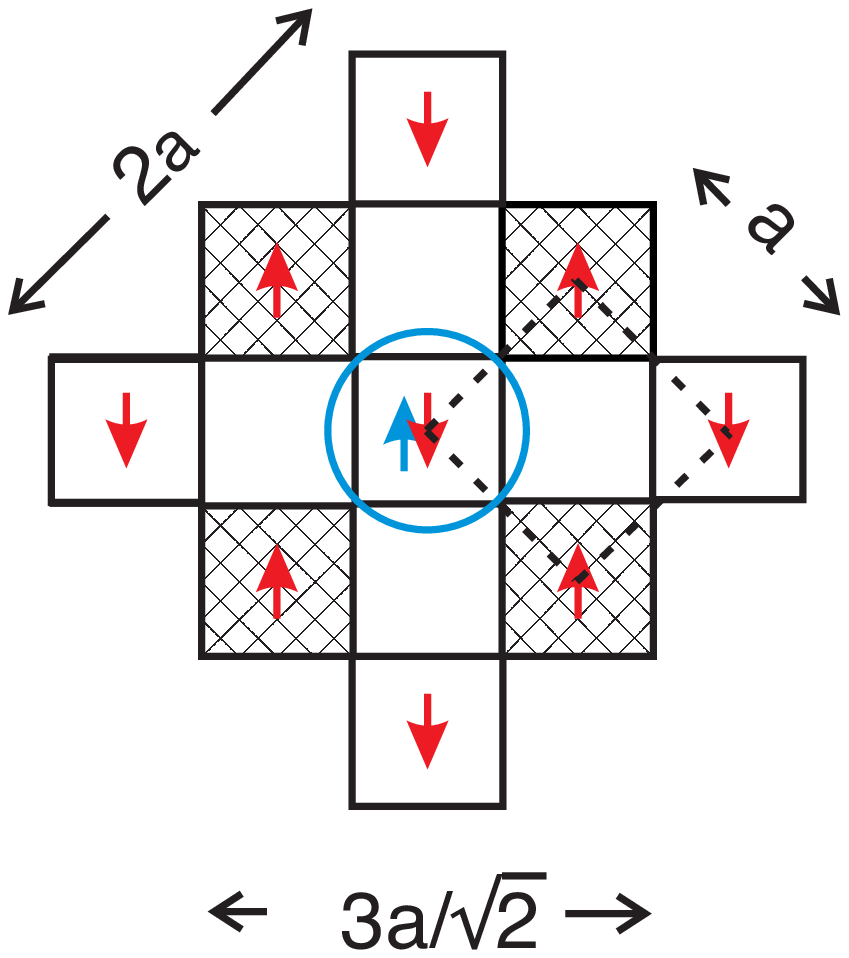}
\caption{\label{sps}Self-protecting singlet (SPS) in an 
antiferromagnetic CuO$_2$ lattice. 
The open circle in the center of the hatched 
supercell of  $3\times 3$ $nn$ 
oxygen--oxygen cells indicates the symmetric oxygen cage accomodating a ZR 
singlet of the copper and the oxygen hole. The cross-hatched squares 
protect the ZR singlet against other holes with their intact spins (``spin 
fence''). The SPS covers an area of $9/2a^2$ in the CuO$_{2}$ lattice.}
%

%
\includegraphics*[width=7cm]{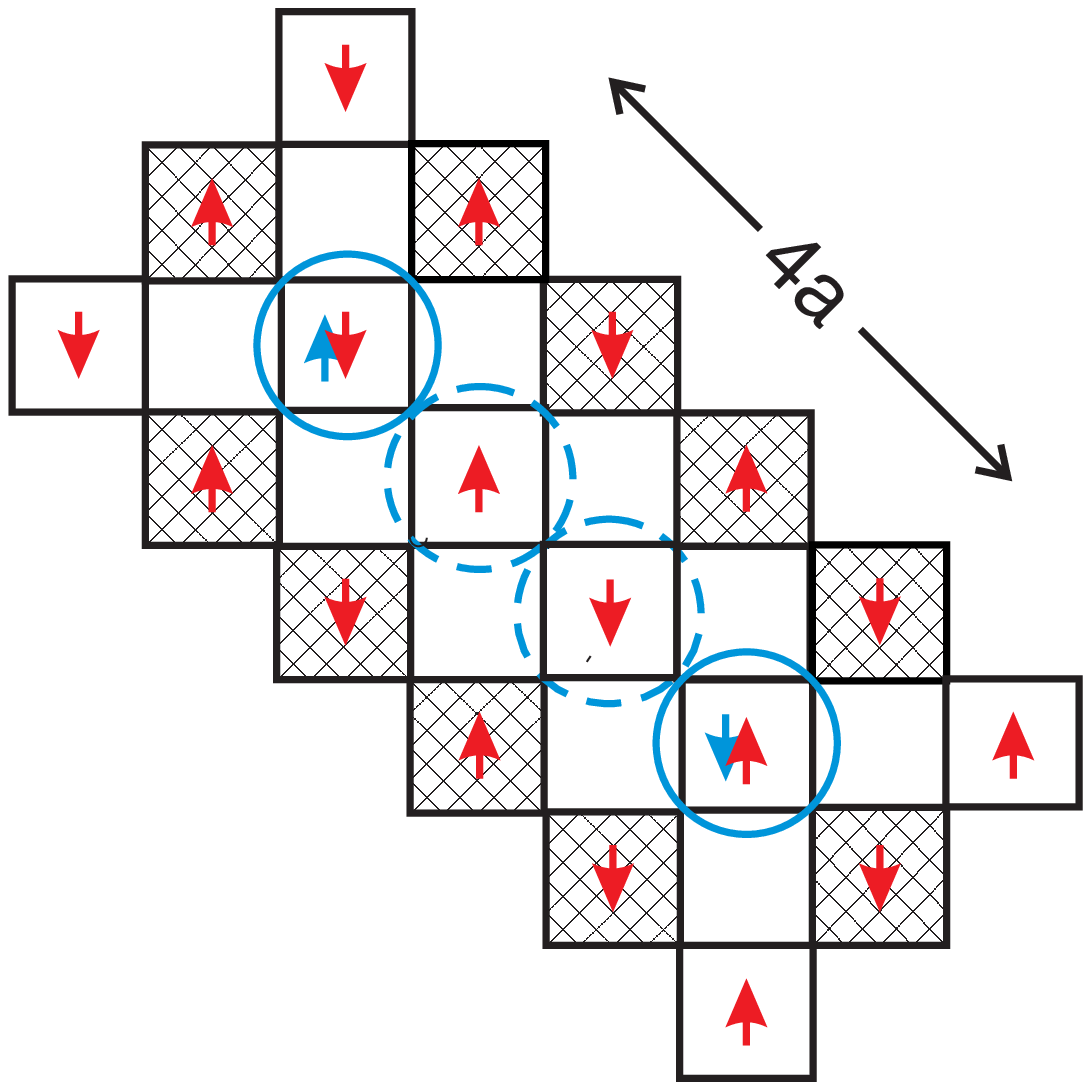}
\caption{\label{psps}Two self-protecting singlets sharing a common 
oxygen site along the Cu--O bond direction create one $paired$ self-protecing 
singlet (PSPS). The spin fence in the cross-hatched squares protects the 
exchange between the two SPS against other holes and the loss of symmetry. 
The PSPS covers an area of $6a^2$ in the CuO$_{2}$ lattice. The PSPS 
extends over four oxygen cages along $a$.}
\end{figure}

Fig.  \ref{psps} displays two self-protecing singlets sharing a common
oxygen site along the direction of the Cu--O bond.  This conformation
may lead to an exchange of holes between two intact ZR singlets
(closed circles) through opened spin fences as indicated by the dashed
circles.  The thus connected SPS are fully antisymmetric in respect to 
the connecting oxygen and may form
in principle a paired ZR singlet state as proposed by M{\"u}ller \cite{Mue03}. 
Self-protection of the ``paired'' singlet will be achieved by an 
enlarged spin fence as indicated by the cross-hatched squares in Fig. 
\ref{psps}.  The area, covered the by the spin fence of the PSPS, 
is nominally $12a^2$, significantly larger than $9a^2$, 
the area protected by the spin fences of two
nonbonding SPS. Thus formation of a PSPS will reduce the area available
to other singlets much more than most closely packed SPS. Notably the PSPS
extends over $4a\simeq 15$ {\AA} (a chain of four corner-linked oxygen 
cages), nearly coinciding with the 
experimentally established planar superconducting coherence length
$\xi_{ab}$.

Because of the the contraction of the Cu--O bonds upon singlet formation both, 
the SPS and the PSPS, are expected to exhibit important vibrational 
properties. Notably the oxygen holes are expected to breathe within 
the rigid $2a$ cell of the spin fence as observed in 
$e.g.$ optimum doped La$_{1.85}$Sr$_{0.15}$CuO$_4$ 
by McQueeney $et$ $al.$ \cite{McQPet}.      

\subsection{Characteristic distributions of SPS and PSPS}

\begin{figure}
\includegraphics*[width=13cm]{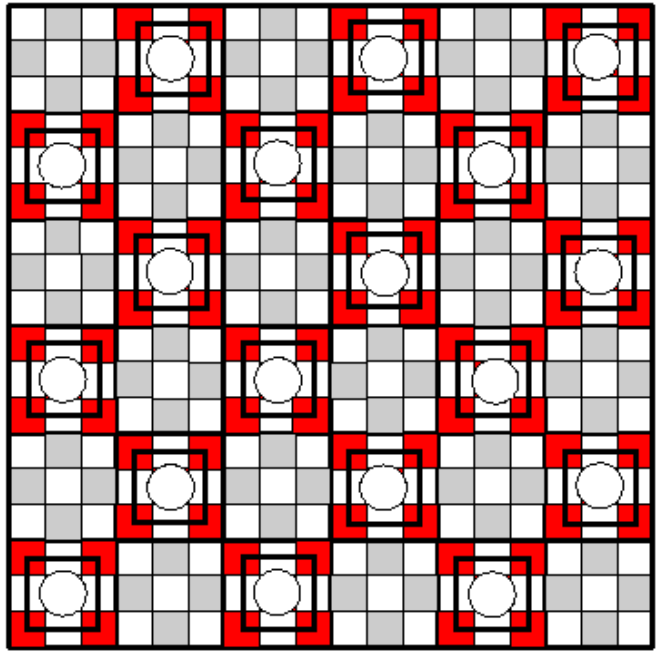}
\caption{\label{ordered}''Stripy'' doping: Perfectly ordered distribution of oxygen holes 
over the antiferromagnetic CuO$_2$ lattice by 
self-protecting singlets indicated by the open circle and the squared 
spin fence. Note the checkerboard lattice of ``crosses'' (dark/red) 
and ``squares'' (grey). Only 0.11 holes/Cu are accomodated.}
\end{figure}

\begin{figure}
\includegraphics*[width=13cm]{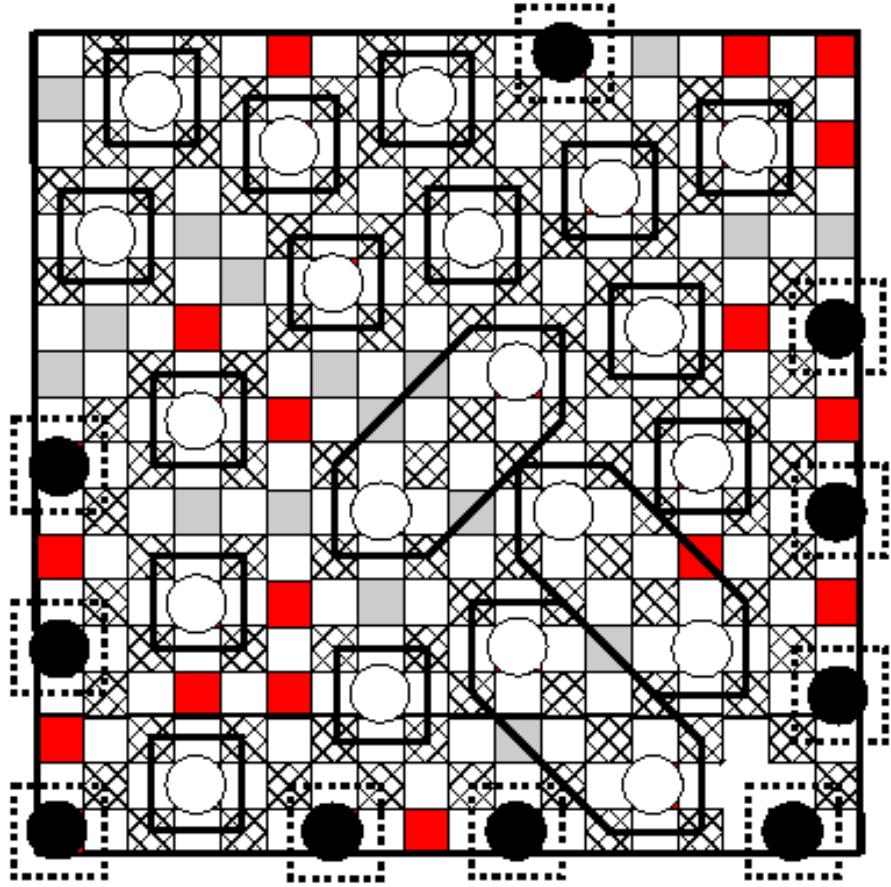}
\caption{\label{critical}Critical doping: Possible high density packing 
of oxygen holes in the antiferromagnetic CuO$_2$ 
lattice by SPS and PSPSs over the ``crosses'' and 
``squares''.  Cross-hatched cells indicate protecting areas. 
Note that the distribution must is inhomogeneous 
favoring the formation of paired singlets (large cigar-shaped spin fences). 
Boundary effects are indicated by filled circles and dotted 
broken spin fences. In this example about 0.18 holes/Cu are 
accomodated.}
\end{figure}

\begin{figure}
\includegraphics*[width=13cm]{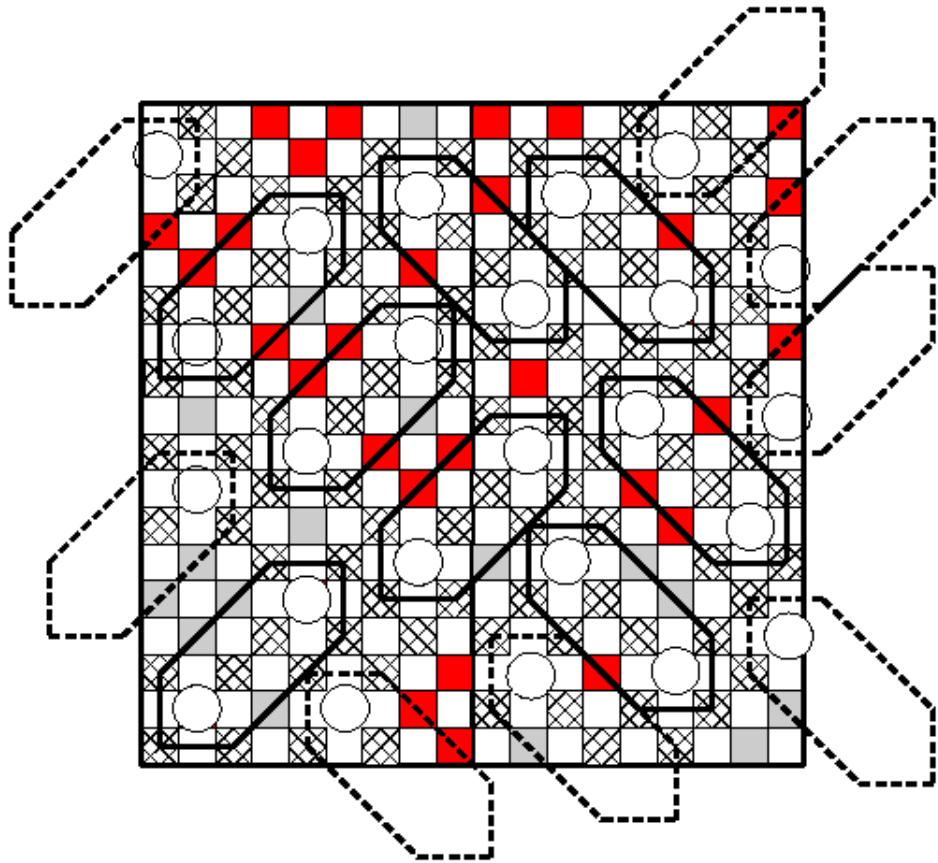}
\caption{\label{optimum}Optimum doping: Nearly closest  packing 
of oxygen holes in the antiferromagnetic CuO$_2$ 
lattice by PSPSs. Note that 
closest packing of PSPSs will pair all
available SPS. In this example about 0.15 holes/Cu are accomodated. 
Boundary effects are indicated by dashed spin fences.}
\end{figure}

\begin{figure}
\includegraphics*[width=13cm]{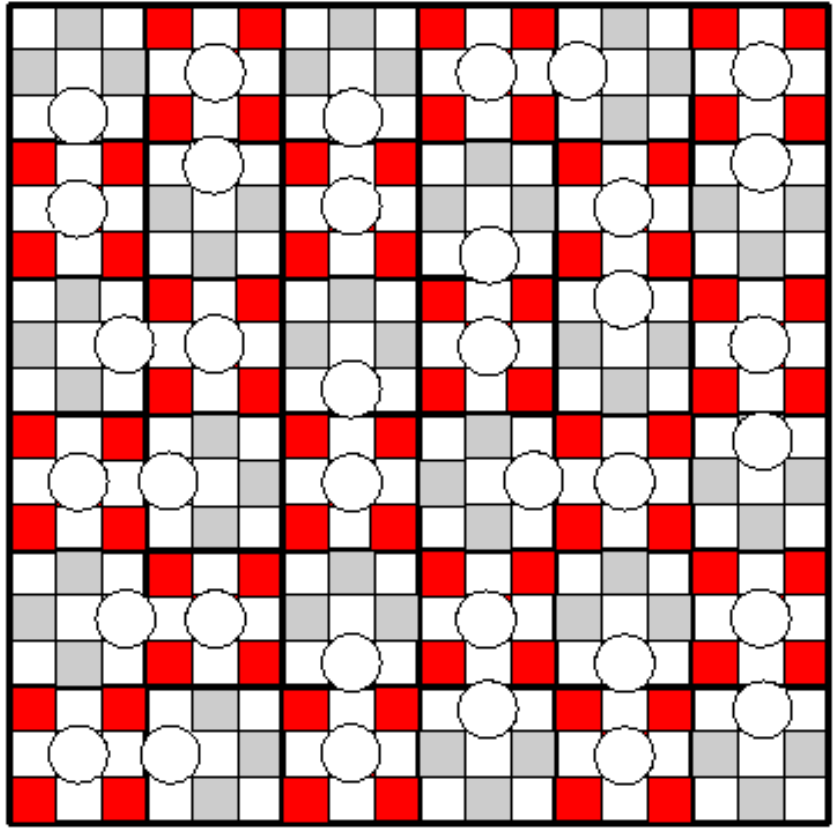}
\caption{\label{over} 
Destructive doping: In this example 0.22 holes/Cu are accomodated. 
All spin fences are broken.}
\end{figure}

Consider the CuO$_2$ lattice as an array of corner linked 
oxygen cages each enclosing one Cu atom (Figs. 
\ref{ordered}-\ref{over}). One SPS covers five of these 
cages and thus forms a $3\times 3$ supercell. The $3\times 3$ supercells 
will create a checkerboard lattice of black (dark/red) fields with  
``crosses'' of 5 cages, and white (grey) fields with ``squares'' of 4 cages. 
Doping will distribute the holes as SPS over this checkerboard 
lattice. At sufficiently high concentrations the SPS will touch each 
other, form clusters, or most likely PSPSs. A full occupation of the lattice 
may be realized by many different distributions 
leading to a remarkable sequence of characteristic hole concentrations. 
Figs. \ref{ordered}-\ref{over} sketch four possible distributions 
relevant for the understanding of the phase diagram of the cuprates. 
The SPS are indicated by objects with a circular cage 
(the singlet hole) enclosed by a square (the spin fence).   

\subsubsection{Stripy long range ordering: $n=0.11\simeq n_s$}
In Fig. \ref{ordered} each ``cross'' (dark/red) is occupied with one SPS. 
Hence the ``squares'' (grey) will not be available for the SPS. The SPS 
will thus align in crossing chains along $a$. One hole rich chain 
alternates with two empty chains creating therewith a $3a$ superstructure. 
Presumably pairing of neighbored SPS to PSPSs will be suppressed by 
long range ordering. Only 0.11 holes/Cu are 
accomodated by this ``stripy'' distribution.

\subsubsection{Critical Doping: $n=0.20\simeq n_{crit}$}
To achieve closer packing of holes the SPS have to be distributed over
the crosses $and$ the squares of the checkerboard lattice.  Fig. 
\ref{critical} displays a possible distribution for a high
density of holes.  Cells excluded by the spin fences are
cross-hatched.  The distribution for the maximum number of
SPS will be realized by most closely packing at $n_{crit}$. 
For $n>n_{cit}$ the spin fences will overlap and destroy the
self-protection of the singlets.  Interestingly this leads generically
to an inhomogeneous hole distribution, apparently because the 
``crosses'' and ``squares'' in the checkerboard lattice are not equivalent for the cross-shaped 
SPS. The inhomogenous hole distribution favors the formation of PSPSs
as it is indicated by the cigar-shaped spin fences enclosing the two
neighbored SPS connected along $a$.  The SPS may only disappear for
dopings exceeding $n_{crit}$. Boundary effects destroying 
the self-protection of the singlets will result from the finite size of 
the crystal. SPS damaged by boundary effects are indicated by filled circles
with dashed spin fences.  About $\simeq 0.18$ holes/Cu may be
accomodated in this distribution. Closest packing of SPS 
accomodates $1/5=0.2$ holes/Cu.

\subsubsection{Optimum Doping; $n\simeq 0.16\simeq n_{opt}$}
Fig.  \ref{optimum} displays as an example a distribution of nearly
closest packed paired self protected singlets.  About 0.15
holes/Cu may be accomodated by this distribution. 
The boundary effects are even more serious for the PSPSs than for
the SPS: PSPSs will be destroyed within $3a$ from the edges of the
crystal. Closest packing of PSPSs accomodates $1/6=0.166$ holes/Cu.

\subsubsection{``Destructive'' doping: $n_{des}> 0.2$}
For $n_{des}= 0.22$ each SPS must overlap its spin fence with
that of the $nn$ SPS (see Fig.  \ref{over}).  At such high hole
concentrations the singlets will be unable to protect against other
holes, break apart or possibly localize.  Phrased in terms of the
classical theory of gases: for $n_{des}$ the antiferromagnetic
lattice may no longer serve as the walls of the container enclosing 
the liquid.

\section{\label{concluding}Summary and Concluding remarks}

The ubiquitous observation of a hump in $B(x)$ of cuprate 
superconductors has led us to an attempt analyzing the quantum liquid 
confined to doped CuO$_2$ lattices in terms of the theory of classical 
real gases or liquids. Most importantly here the walls of the container enclosing the
quantum liquid are formed by the antiferromagnetic lattice which 
couples to the atomic lattice by creation of self-protecting spin singlets. 
It is basically internal magnetostriction, although of a novel type, 
that allows for insights into the structure of strongly correlated 
electrons from an analysis of the lattice parameters. 

We have shown that the hump is strong evidence
for the existence of relatively large incompressible particles
interacting with each other.  The area of the particles can be
estimated from the location of the hump around $x_{opt}=0.16$, and 
from its collapse around $x_{crit}>x_{opt}$, respectively. At a microscopic level
neither simple ($nn$) singlets, nor the ZR-singlet will match the
area of $6.25 a^2 $ to be covered for hard core repulsion at
$n_{opt}= 0.16$.  The area of pairs of PSPS
however is close to this requirement.  On the other hand the area $9/2
a^2$ covered by an unpaired SPS leads to close packing at $n_{crit}\simeq 
0.20$, and seems to match even the number of destructive 
doping $n_{des}\simeq 0.22$ where in IR experiments the 
bosonic excitations disappear \cite {Tim03}. 

The SPSs are most likely bosonic, 
and not fermionic. Clearly this and other issues of the SPS and the 
PSPS will raise a bunch of challenging questions, which certainly cannot be 
answered from our so far almost only phenomenological 
approach which lead us via a geometrical construction to 
these particles.
It is clear however that displacive lattice degrees of freedom must be 
involved in the formation of the singlet states. Otherwise the hump 
would not occur.

\begin{acknowledgments}
I appreciate stimulating and enlightening discussions with K. A. M{\"u}ller, T. Egami 
and E. Stoll during this conference. 
    
\end{acknowledgments}

\bibliography{Bled2003}

\newpage


%

%

%

%

\end{document}